\documentclass{sigchi}

\usepackage{balance}  % to better equalize the last page
\usepackage{graphicx} % for EPS, load graphicx instead
\usepackage{times}    % comment if you want LaTeX's default font
\usepackage{url}      % llt: nicely formatted URLs
\usepackage{todonotes}

\makeatletter
\def\url@leostyle{%
  \@ifundefined{selectfont}{\def\UrlFont{\sf}}{\def\UrlFont{\small\bf\ttfamily}}}
\makeatother
\def\pprw{8.5in}
\def\pprh{11in}

\usepackage[pdftex]{hyperref}
\hypersetup{
pdftitle={SIGCHI Conference Proceedings Format},
pdfauthor={LaTeX},
pdfkeywords={SIGCHI, proceedings, archival format},
bookmarksnumbered,
pdfstartview={FitH},
colorlinks,
citecolor=black,
filecolor=black,
linkcolor=black,
urlcolor=black,
breaklinks=true,
}

\begin{document}

\title{Towards Around-Device Interaction using Corneal Imaging}

\numberofauthors{2}
\author{
  \alignauthor Daniel Schneider\\
    \affaddr{Coburg University}\\
    \email{daniel.schneider@hs-coburg.de}
  \alignauthor Jens Grubert\\
    \affaddr{Coburg University}\\
   \email{jg@jensgrubert.de}\\
}
%\author{Daniel Schneider\thanks{e-mail: daniel.schneider@hs-coburg.de} %
%\and Jens Grubert \thanks{e-mail: jg@jensgrubert.de}} %
%%\and Martha Stewart\thanks{e-mail:martha.stewart@marthastewart.com}}
%\affiliation{\scriptsize Coburg University}

%\author{
 % \alignauthor Daniel Schneider\\
    %\affaddr{Coburg University}\\
    %\affaddr{Friedrich-Streib-Str. 2}\\
    %\affaddr{96450 Coburg}\\
   % \email{daniel.schneider@hs-coburg.de}\\
  %  \affaddr{+49 9651 317 354}
 % \alignauthor Jens Grubert\\
%    \affaddr{Coburg University}\\
    %\affaddr{Friedrich-Streib-Str. 2}\\
   % \affaddr{96450 Coburg}\\
  %  \email{jg@jensgrubert.de}\\
 %   \affaddr{+49 9651 317 279}   
%}

%\author{
%  \alignauthor anonymous\\    
%  \alignauthor anonymous\\   
%}

\maketitle

\begin{abstract}
Around-device interaction techniques aim at extending the input space using various sensing modalities on mobile and wearable devices. In this paper, we present our work towards extending the input area of mobile devices using front-facing device-centered cameras that capture reflections in the human eye. As current generation mobile devices lack high resolution front-facing cameras we study the feasibility of around-device interaction using corneal reflective imaging based on a high resolution camera. We present a workflow, a technical prototype and an evaluation, including a migration path from high resolution to low resolution imagers. Our study indicates, that under optimal conditions a spatial sensing resolution of 5 cm in the vicinity of a mobile phone is possible.
\end{abstract}

\keywords{
around-device interaction; camera-based interaction; corneal imaging
%	Guides; instructions; author's kit; conference publications;
%	keywords should be separated by a semi-colon. \newline
%	\textcolor{red}{Optional section to be included in your final version, 
  %but strongly encouraged.}
}

\category{H.5.2.}{User Interfaces}{Input devices and strategies}

%See: \url{http://www.acm.org/about/class/1998/}
%for more information and the full list of ACM classifiers
%and descriptors. \newline

%\textcolor{red}{Optional section to be included in your final version, 
%but strongly encouraged. On the submission page only the classifiers’ 
%letter-number combination will need to be entered.}

\section{Introduction}

Since over ten years, smartphones are a popular interaction medium, as they allow us to interact in a multitude of mobile contexts. However, small device sizes, aiming at increased mobility~\cite{729537},  can sacrifice the input space of those devices. If devices shrink, while fingers stay the same, interaction may become inefficient~\cite{FatFinger}. Hence, there is a need for compensating for the lack of physical interaction area. 

One option is to extend the input space of interactive displays using sensors, leading to a decoupling between input and output space \cite{harrison2010appropriated}. Various research has sparked in the area of around-device interaction,  extending the input space to near-by surfaces or to mid-air. So far, most research focused on equipping either mobiles~\cite{butler2008sidesight}, the environment~\cite{radle2014huddlelamp} or the user~\cite{chan2015cyclops,grubert2015multifi, grubert2016glasshands} with additional sensors. 
However, deployment of such  hardware modifications is hard. Market size considerations discourage application developers, which limits technology acceptance in the real-world~\cite{buxton2008long}. 
%Only few research aims at extending the input around unmodified mobile devices (e.g., \cite{nandakumar2016fingerio}).

We envision a future in which \textit{unmodified} mobile and wearable devices that are equipped with standard front-facing cameras allow for ample movements, including the environment around and to the sides of the device, without the need for equipping the device or the environment with additional sensing hardware. This could be achieved by utilizing the combination of built-in camera and the reflection in the human eye as lens-mirror (or catadioptric) system. Specifically, the environment of the user is reflected in the eye, which in turn can be captured by the device camera. However, to date, front-facing cameras typically do not posess the resolution needed for corneal imaging at typical interaction distances of handheld devices. Hence, in this paper, we investigate the feasibility of employing corneal imaging techniques with a mobile device-centred high-resolution camera. We present a pipeline for corneal imaging targeted at interaction with mobile devices and  a user study investigating the performance of our approach under laboratory conditions. Besides looking at possible resolutions of future cameras, we also investigate a possible migration path up to resolution available in current mobile device cameras.

\section{Related Work}\label{sec:relatedWork}
%\vspace{-0.05cm}
%In this section we present and overview of related work in the fields of corneal imaging and around-device interaction.
%In this section we present and overview of related work in the fields of corneal imaging system, gaze estimation, interaction with mobile  devices and around-the-body interaction, which inspired this paper. It follows an overview of the related work in the listed fields.

%The direct predecessor of this paper is GlassHands from Grubert et al. \cite{grubert2016glasshands}. 
\subsection{Corneal Imaging}
Nishino and Nayar \cite{nishino2006corneal} used the term "corneal imaging system" to describe as catadioptric (mirror + lens) system with combination of the cornea of an eye, as mirror, and the camera capturing the appearance of the eye. Nishino and Nayar give a comprehensive analyses of the information about environment embedded in a single image of the eye. In this context, they describe several characteristics of a catadioptric system with cornea, as mirror, and camera. They also describe how to reconstruct a 3D object using both eyes. In contrast to the reconstruction using both eyes, we will only use a single eye for reconstruction in our evaluation to cover a wider interaction space. Nitschke et al. \cite{nitschke2013corneal} give an overview of applications, which use corneal imaging in various settings. One uses the reflection to get information about the illumination of an Image \cite{nishino2004eyes}. Another possible application is espionage. With corneal imaging an image with the eye can show the content of screen a person looks at \cite{backes2009tempest}. Further, a person's  gaze direction and gaze point can be estimated \cite{nitschke2013see}. % More information about gaze estimation can be found in the next subsection.  
 Nitschke et al. \cite{nitschke2011display} uses multiple images with various gaze directions to reconstruct an object and Schnieders et al. \cite{schnieders2010reconstruction} uses the gaze direction of both eyes for the reconstruction. Typically, corneal imaging system focus on head-mounted eyet tracker, but applications for gaze estimation also have been proposed for mobile (e.g., \cite{Khamis:2016:GMA:2851581.2892314}) and public displays (e.g., \cite{Khamis:2016:CDS:2968219.2968342}) as well as smartwatches (e.g., \cite{esteves2015orbits}). % As mentioned above we use only a single image of one eye for the reconstruction. Other application domains include optical see-through head-mounted display calibration~\cite{plopski2015corneal} or object recognition~\cite{takemura2013estimating}.

%\subsection{Gaze Estimation}
%For the estimation of the gaze direction it is often assumed that the gaze direction can be approximated  to the orientation of the eye (for more information about orientation of the eye and gaze direction see section \ref{sec:basicEyeModel}). The orientation of the eye is on part of the reconstruction process described in section \ref{sec:design}. There are various algorithm for gaze estimation. It can be distinguished between passive methods \cite{wang2001gaze, nishino2006corneal, nitschke2011display, schnieders2010reconstruction, wood2014eyetab, jariwala2016robust} that work on any eye image  and active methods that require additional controlled illumination \cite{nakazawa2017eye, guestrin2006general, villanueva2007models}.

%For the eye orientation we use a passive method based on the algorithm of Nishino and Nayar \cite{nishino2006corneal}, Nitschke et al. \cite{nitschke2013corneal} and Wood et al. \cite{wood2014eyetab}. 

\begin{figure*} [tb]
	\begin{center}
		\includegraphics[width=\linewidth]{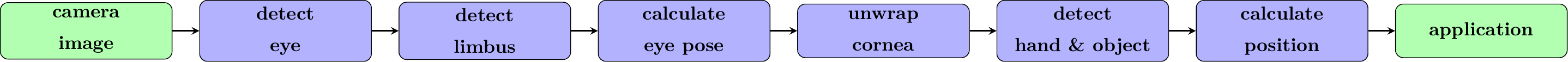}
		\caption{Pipeline for calculation of positions (purple) with input and output (green).}
		\label{fig:pipeline}
	\end{center}
\end{figure*}
%\vspace{-0.5cm}

\subsection{Around-device Interaction}

Along with the reduction of the size and weight of mobile and wearable devices, the need for complementary interaction methods evolved. While the available space on a mobile device is continuously shrinking, research began investigating options for interaction next to~\cite{Oakley:2014:IEO:2556288.2557138}, 
above~\cite{kratz2009hoverflow, Freeman:2014:TUA:2628363.2634215}, 
behind~\cite{DeLuca:2013:BAS:2470654.2481330, Wigdor:2007:LTS:1294211.1294259}, 
across~\cite{Schmidt:2012:CIS:2317956.2318005,chen2014duet}, or 
around~\cite{Zhao:2014:SDI:2642918.2647380,xiao2014toffee} 
the device. The additional modalities are either substituting or complementing the devices' capabilities. %Extending the input space of mobiles and wearables around them was mainly studied on surfaces and in free space, but 
These approaches rely on modifying existing devices using a variety of sensing techniques, which can limit their deployment to mass audiences.

Recently, researchers started to investigate the use of unmodified devices. Nandakumar et al. \cite{nandakumar2016fingerio} propose to use the internal microphones of mobiles to determine the location of finger movements on surfaces, but can not support mid-air interaction. Song et. al \cite{song2014air} enabled in-air gestures using the front and back facing cameras of unmodified mobile devices.  However, their interaction space is limited to the field-of-view of the cameras, constraining the interaction space to two narrow cones in front and behind a device. Much of the interaction space around the mobile device, such as the areas to the sides of the device are not observed by these cameras. In contrast, our work focuses on a larger interaction space, covering positions around the device sides, as well as large hand gestures.
With Surround See, Yang et al.~\cite{yang2013surround} modified the front-facing camera of a mobile phone with an omnidirectional lens, extending its field of view to $360^\circ$ horizontally. They showcased different application areas, including peripheral environment, object and activity detection, including hand gestures and pointing, but did not comment on the recognition accuracy.

The closest work to ours is GlassHands \cite{grubert2016glasshands}. Grubert et al. demonstrated how the input space around an unmodified device can be extended, by using a built-in front-facing camera of an unmodified handheld device and some reflective glasses, like sunglasses, ski goggles or visors. The approach of Grubert et al. is to locate the hand and the handheld device in the reflection of the glasses and calculate their relative position in a common coordinate frame, hence virtually extending the touch screen of the mobile device. However,  the usage of dark glasses in low-light environments may not be appropriated due to perceptual and social reasons. In this work, we propose to modify their original idea to integrate corneal imaging techniques. In our work, the replacement of the mirror (from glasses to the cornea) is the first major difference. Due to different forms of the reflection area the approach to detect these are also different (Grubert et al used a simple thresholding approach for detecting the glass area). %Grubert et al. next analyze the scene and then make the unwrapping of the relevant coordinates of the detected objects only. We make the unwrapping of the complete picture and then analyze the scene. 
For  unwrapping Grubert et al. proposed a special calibration target and process for unwrapping images from curved glasses. The cornea model we use makes this calibration step redundant.

\section{Concept and Implementation}\label{sec:design}
Our approach aims at detecting the position of the user's hand relative to a mobile handheld device such as a smartphone. This way, the hand can be used as input device in the vicinity of the device.  To calculate the position of those objects of interest in a reflected area, one needs to be able to detect the reflection area, remove the distortion from the shape of the cornea, detect objects (in our case a handheld device and a hand), and, finally, reconstruct the relative position of those objects in a common coordinate frame. The workflow is summarized in Figure \ref{fig:pipeline}. First, an input image is searched for an eye, then the limbus, the contour of the cornea, and, therefore, the contour of the reflected area, is detected and the eye pose and location relative to the camera is calculated. Next, this information is used for unwrapping the the corneal region, and, afterwards, analyze the scene based on the resulting image. In our case, we search for objects of interest (such as a hand or a finger) in the vicinity of a smartphone and determine their relative positions in a common coordinate frame. This data is finally passed to an application which makes use of it. 

\begin{figure*} []
	\begin{center}
		\includegraphics[width=\linewidth]{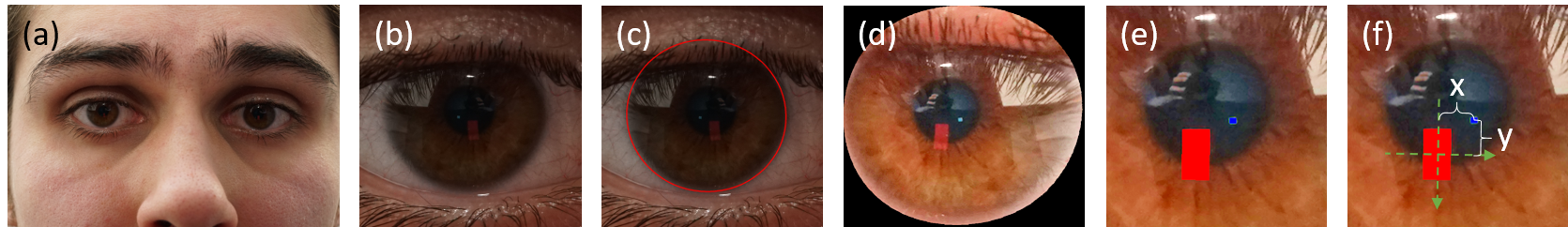}
		\caption{Results of single steps of the pipeline, (a) original camera picture, (b) eye region, (c) eye region with limbus, (d) Unwrapping result, (e) Object detection (red and blue rectangle) (f) distance of objects.}
		\label{fig:pipelineResults}
	\end{center}
\end{figure*}

Figure \ref{fig:pipelineResults} provides an overview of the process data. The content of the pictures are related to the laboratory study presented below. 
%The scene (fig. \ref{fig:pipelineResults} (a) gives an impression on the setup of the scene. 
The input data (original image) is shown in Fig. \ref{fig:pipelineResults} (a). Fig. \ref{fig:pipelineResults} (b) shows the eye area for the right eye and is the result of the eye detection. The result of the limbus detection is depicted in Fig. \ref{fig:pipelineResults} (c). In Fig. \ref{fig:pipelineResults} (d), the distortion of the cornea is partial compensated during the unwrapping step. Objects of interest Fig.  \ref{fig:pipelineResults} (e) are detected and their relative distance is calculated (Fig.  \ref{fig:pipelineResults} (f)).

The first step of the pipeline is the detection of the eyes.  For automatic eye detection the approaches of Kothari and Mitchell \cite{kothari1996detection} or Kawaguchi et al. \cite{kawaguchi2005detection} can be used. Please note, that these approaches assume a full face to be visible in the image. 

% \subsection{Eye Detection}
% The first step of the pipeline is the detection of the eyes. For each eye the eye region is determined separately. For the transformation from Cartesian to polar coordinates during limbus detection square regions are necessary. Therefor the eye detection returns square regions.

% There are different approaches for the estimation of the eye region in an Image. Kothari and Mitchell \cite {kothari1996detection} locate the position based on a series of images. In a gray scale images the pattern of the dark iris sorunded by the bright sclera is located. This are potential candidates. The position that has the most candidates over all images is the eye region.
% Another approach locates potential circles and compares them with a template to get the best candidate. \cite {kawaguchi2005detection}.

% We select the eye image manually.

%\subsection{Limbus Detection}
%The next step of the the pipeline is the limbus detection. We are using a RANSAC-based algorithm \cite{fischler1981random} by Wood and Bulling \cite{wood2014eyetab} to detect the limbus, as an ellipse in the image.
The next step of the the pipeline is the limbus detection. We are using a RANSAC-based algorithm %\cite{} 
by Wood and Bulling \cite{wood2014eyetab} to detect the limbus as an ellipse in the image. The detected limbus is improved by predefined sample-sets of the RANSAC algorithm.
% For the limbus detection the algorithm of Wood and Bulling \cite{wood2014eyetab} is used. 

% At the beginning, the eye region is scaled to a size of 150x150 pixels, in order to reduce the data to be processed in high-resolution images and to provide a output with constant for further processing. Subsequently, the image is transferred from Cartesian coordinates in to polar coordinates, referred to as a polar image. The origin for the Cartesian coordinate system lies in the center of the eye region. In the eye region it is assumed that the center of the pupil is close to that origin. A possible misclassification of the iris contour by the pupil or eyelid is limited by a lower and an upper limit. The edge detection on the polar image is used to find potential limbus points. The points from the polar image are then transformed back into the original coordinate system. With an RANSAC algorithm \cite{fischler1981random} is used to select the best ellipse. The quality (best) of an ellipse is made on the basis of the number of potential points which lie on or just next to the ellipse.

To calculate the pose of the eye, we use an eye model of two intersecting spheres from Nitschke et al.~\cite{nitschke2013corneal}. For the corneal reflection, the sphere of the model containing the cornea is used. One pole of the sphere lies at the center of the pupil. Alternatively, more advanced eye models can be used (e.g., \cite{wood20163d}). As proposed by Nishino and Nayar \cite{nishino2006corneal}, the position and orientation of the eye can be calculated with the limbus center. First, the distance between the camera center of projection and the center of the limbus has to be determined.  Under the assumption of weak perspective projection, the limbus in 3D space is ﬁrst orthographically projected onto the average depth plane, which is parallel to the image plane and passes through the center of the limbus. %Since the limbus is a circle in 3D space, this average depth plane always passes through the center of the limbus. 
The limbus in the projection plane is an ellipse. With the orientation of the ellipse and the ratio of major and minor axis, representing the orientation of the pol-axis, the center of the sphere can be reconstructed relative to the camera.

With the eye pose and the location of the eye, the pixel of the reflection in the image can be projected back to the surface of the mirror (cornea). In order to process the surface of the cornea further, the texture is unwrapped. For the unwrapping, we use a equirectangular projection \cite{snyder1997flattening}. This type of projection has higher distortions in the polar region than around the equator. To minimize distortions, we rotate our eye model 90\(^{\circ}\) around the x-axis, so the cornea is no longer the polar region of the. These rotation must be taken into account and reversed during further processing.

Then, scene analysis algorithms can be applied to this unwrapped image.We detect the smartphone display and the fingers in a similar manner as in GlassHands \cite{grubert2016glasshands}. For interactive purposes, we aim at reconstruction the relative positions of the display and the hand in a common coordinate frame (the display plance of the mobile device, virtually extended around the device). We assume that the interaction takes place in this virtual plane.
% as  (under the assumption

%\vspace{-0.5cm}

% \subsection{Unwrapping}
% To detect objects on the surface of a sphere the texture has to be unwrapped. One way to do so is the equirectangular projection. With this projection the surface of a sphere can be mapped to a plane. Therefor the angles of the polar coordinates are mapped to the x- (\(\phi\)) and y-axis (\(\tau\)) of the texture. 

% The equirectangular projection creates distortion at the pole of a sphere, where in the case of the eye model the iris and pupil is. This produces highly distorted Images like \ref{fig:ArtefaktAbwicklungPol} (c). With a rotation around 90\(^{\circ}\) such artifacts can be avoided. This rotation must be taken into account and reversed during further processing. 
% \begin{figure}[!t]
% 	\centering
% 	\includegraphics[width=\linewidth]{RotationBackProjection.png}
% 	\caption{Artifact equirectangular projection at pole. (a) input with limbus (red), (b) projection with rotation, (c) projection without rotation }
% 	\label{fig:ArtefaktAbwicklungPol}
% \end{figure}

%\subsection{Hand and Object Detection}
%For the detection of the mobile device it is assumed it is a single color rectangle. For the detection of this object the texture is transformed into a hsv color space. In this new image, the pixels with the correct color are selected via threshold. In the binary image the objects are located and the best candidate of the object will be selected. The best candidate is the biggest rectangle, or nearly a rectangle, nearest to the center of the texture.

\begin{figure} [!b]
	\begin{center}
		\includegraphics[width=0.4\columnwidth]{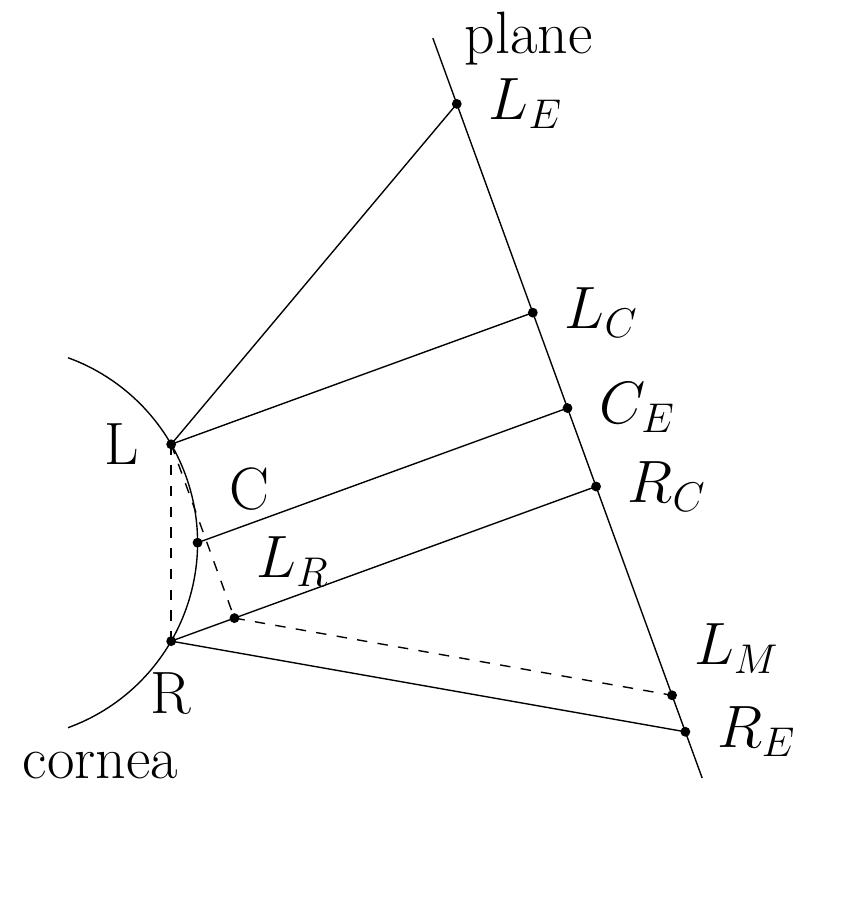}
		\caption{Schematic view from cornea and reconstruction plane. Mobile device on cornea has the left (L), center (C) and right point (R). Mobile device on plane has the left (\(L_E\)), center (\(C_E\)) and right point (\(C_E\)). Projection with \(\vec{CC_E}\) from L to \(L_C\) and R to \(R_C\) on plane. \(L_R\) and \(L_M\) are mirrored points of L and \(L_E\) on line C \(C_E\)}.
		\label{fig:reconstruction}
	\end{center}
\end{figure}
%\vspace{-0.3cm}
%\subsection{3D Reconstruction and positioning}

For the reconstruction of the plane with mobile device and hand, it is assumed that the mobile device can be approximated by a rectangular shape and the projection rays from the pinhole of the camera are parallel. The first step in the reconstruction, is the calculation of the left (L), center (C) and right (R) points of the object on the cornea are needed. With the reflection, these points are projected to the points \(L_E\), \(C_E\) and \(R_E\) on the plane. The width of the real object is the distance between \(L_E\) and \(R_E\). Using the distance between L and R, the distance between \(L_C\) and \(R_C\) can be calculated. The same applies for the point \(L_R\) and the distance from this to R. The distance between \(R_C\) and \(L_M\) is the same as \(L_E\) and \(L_C\). With the ratio of the distance between \(R_C\) and \(L_M\) and the distance between \(R_C\) and \(R_E\) the distance to the plane can be calculated.

To calculate the orientation of the plane, the ratio of width and height between the image and physical dimensions are used. The center of the mobile device is the origin in the object plane. The coordinates of the hand will be calculated via intersection of the reflection vector and the plane, and, then, transformed into plane coordinates. 

If a metric distance is not needed between smartphone and hand  (e.g. for spatial gestures), one could, alternatively, determine the relative distances in pixel space, and, potentially, skip the unwrapping and reconstruction steps. 

The pipeline was implemented in C++ and OpenCV was used for image processing functions.
% - calculate L and R
% - calculate distance:
%   - assumption: rays from pinhole to cornea are parallel -> same angle vrvc vcvl

% For the reconstruction of the plane with mobile device and hand, it is assumed that the object of the mobile device is a rectangle. For the reconstruction the ratio of width in the texture and the real width is calculated. With this factor the distance of the plane can be calculated. To calculate the orientation of the plane the ratio of width and height in image and in real are used. The center of the mobile device is the origin in the object plane. The coordinates of the hand will be calculated via intersection of the reflection vector and the plane and then transformed into plane coordinates.
%\section{Implementation}\label{sec:implementation}
%\lipsum

\section{Evaluation}\label{sec:evaluation}
To determine the accuracy of the position estimation of the algorithm, and, thus, to answer the question whether it is suitable for the interaction with the mobile device, a laboratory study was carried out. In this study, a highly controlled environment is selected to present the usage under optimal conditions. %Especially the lightning is some aspect which often can not be found in the real world.

\subsection{Design and Apparatus}
\begin{figure} [!t]
	\begin{center}
		\includegraphics[width=\linewidth]{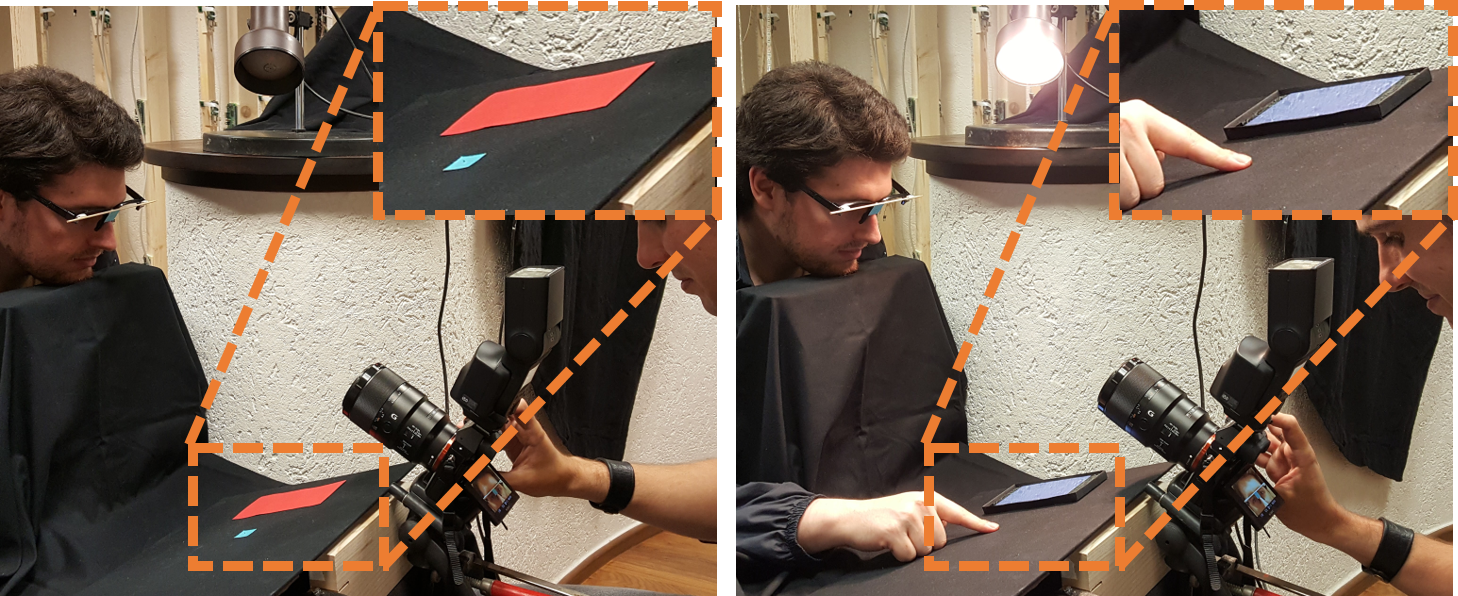}
		\caption{Setup of the laboratory study. Left: Condition \textit{RECT} with red rectangular and blue marker. Right: Condition \textit{FINGER} with smartphone and finger.}
		\label{fig:evaluationSetup}
	\end{center}
\end{figure}

We aimed for descriptive measures of our approach, both under optimal and degraded conditions. To this end, we varied the image resolution and used two detection scenarios (see Figure \ref{fig:evaluationSetup}). The focus in the first scenario (condition \textit{RECT}) was the position estimation for reflected objects under optimal conditions. To this end, we opted for easily detectable colored shapes. The objects used in the study are a red rectangle (7 cm by  14 cm), which represents a mobile device and a blue square (edge length of 2 cm) as a replacement for a finger (see Figure \ref{fig:evaluationSetup}, left). During the study, the distances between the midpoints of the two objects were measured. The distance was divided into an X and a Y component. The X axis is parallel to the narrow edge (7 cm edge) of the rectangle and the Y axis is parallel to the large edge (14 cm edge). To further optimize object detection, a black background was used (high contrast to red and blue object). For a better lightning the flash of the camera was used. 

While still under laboratory conditions, the second scenario (condition \textit{FINGER}) used actual objects that could be used in natural contexts (see Figure \ref{fig:evaluationSetup}, right). These were a Sony Z3 (display size of 6.4 cm by 11.4 cm, black case) showing a white image (following the assumption of Grubert et al. that often the mobile device screen can be the brightest part in a scene \cite{grubert2016glasshands}) and a fist with straightened index finger, which was illuminated by an external light to counterbalance shadows from a chin rest used in the study. During the study, the distance between the smartphone center and the top of the index finger were measured (replicating scenario 1). %Equally to the first part of the study, the distance was devided into an X and a Y component. The X axis is parallel to the narrow edge (6.4 cm edge) of the phone and the Y axis is parallel to the large edge (11.4 cm edge). The background was black. to illuminate the setup an external spotlight was used

%\vspace{-0.3cm}

We envision future built-in cameras to have high resolution imagers (or alternatively, dual wide angle and telephoto lenses as in the Samsung Galaxy Note 8), and, hence, did not want to limit ourselves by today's available front-facing smartphone cameras. Instead, for the recordings, a Sony \(\ alpha \) 7r camera with 36 megapixel sensor and macro lens Sony FE 90mm f / 2.8 Macro G OSS was used. The camera was placed such that the nodal point would coincide approximately with the position of a possible built-in camera (slightly above to still allow object detection).% 10 cm above the red rectangle.

We invited ten people (2 female and 8 male, mean age of 25 years, sd=2.4, variety of iris textures) to participate in the study. After an introduction phase, participants were asked to position their chin on a pedestal at ca. 45 cm distance (mimicking a typical interaction distance for handheld interaction \cite{bababekova2011font}), put on an empty glass frame with a marker and were instructed to look into the camera. %Then, the distance between eyeball and marker on the glass frame was measured. 

The picture taken with the camera captures parts of the face (see Figure \ref{fig:pipelineResults} (a)), and, therefore, the exisiting algorithms (\cite{kawaguchi2005detection, kothari1996detection}) mentioned above can not be used for automatic eye detection in our study setup. Instead, we first use the marker to roughly determine the eye region and then detect and center the real eye region by locating the pupil via thresholding, contour detection and ellipse fitting. 
%The marker was used to quickly determine the eye region of interest in the laboratory condition (which would be replaced with an automated eye detection approach outside of laboratory conditions  \cite{kawaguchi2005detection}).
During the data acquisition, the subjects were asked to keep still and not to change their position. If the participants moved, the data was discarded and positioning and measurements were repeated. In the first part, the blue square and in the second part the participants fist and index finger were positioned at nine different locations. The positions are divided into three columns and three lines to the right of the red rectangle. These are parallel to the X and Y axes. The columns are at a distance 10 cm, 20 cm and 30 cm from the center, and the rows are offset 10 cm up, 0 cm and 10 cm downwards. Four images were taken for each position. After completing the recording, the participants were thanked for their participation in the test. The test setup for both condition are shown in Figure \ref {fig:evaluationSetup}. With these setup there are a total of 360 samples (10 participants x 9 locations x 4 repetitions) for each scenario.

Currently, front cameras in mobile phones do not provide resolutions of our employed camera system. To provide a migration path between high and low resolution imagers, the algorithm is applied to images of different sizes. For this purpose, the height and width of the images are scaled by a factor (0.5, 0.25, 0.125). The camera produces images with a resolution of 7360x4912 px. In these images the eye region is 1000x1000 px. The eye region for factor 0.5 is 500x500 px, for 0.25 250x250px and for 0.125 125x125 px, a resolution achievable with today's cameras. For example, a 2 megapixel smartphone camera in 20 cm distance or a 8 megapixel smartphone camera in 40 cm distance, with wide angle lenses,  produce images with eye regions of ca. 125x125 px.
\vspace{-0.2cm}
%The different sizes of the images is listed in table \ref{table:resolutions}

% \begin{table}[!t]
% 	\begin{tabular}{|c| c | c | c | c |}
% 		\hline
% 		scale:  & 100\% & 50\% & 25\% & 12.5\% \\ \hline
% 		resolution:   & 7360 x 4912 & 3680 x 2456 & 1840 x 1228 & 920 x 614\\ \hline
% 		eye region: & 1000 x 1000 & 500 x 500 & 250 x 250 & 125 x 125 \\ \hline
% 	\end{tabular}
% 	\caption{tested resolutions in pixel}
% 	\label{table:resolutions}
% \end{table}

\subsection{Results}

%\todo[inline]{calculate NHST (t-test) between both conditions}
\begin{figure*}[!t]
	\centering
	\includegraphics[width=0.9\textwidth]{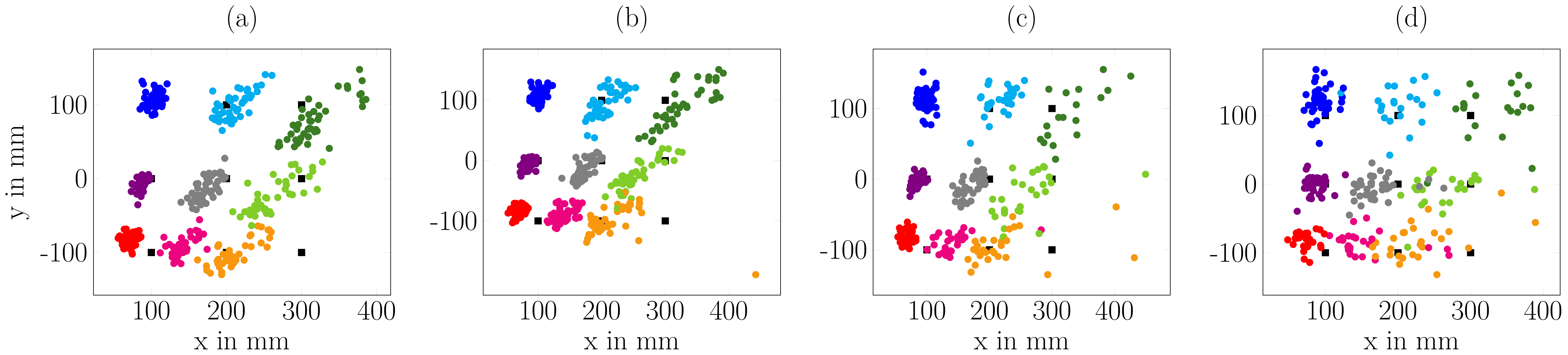}
	\caption{Scatter plot for all resolutions of eye regions of condition \textit{RECT}. Resolution of eye region is (a) 1000 x 1000,  (b) 500 x500, (c) 250 x 250, (d) 125 x 125. The color marks the aimed positions (x, y): red (100, -100), magenta (200, -100), orange (300, -100), violet (100, 0), gray (200, 0) light green (300, 0), blue (100, 100), light blue (200, 100) and green (300, 100).}
	\label{fig:scatterScenario1}
\end{figure*}
\begin{figure*}[!t]
	\centering
	\includegraphics[width=0.9\textwidth]{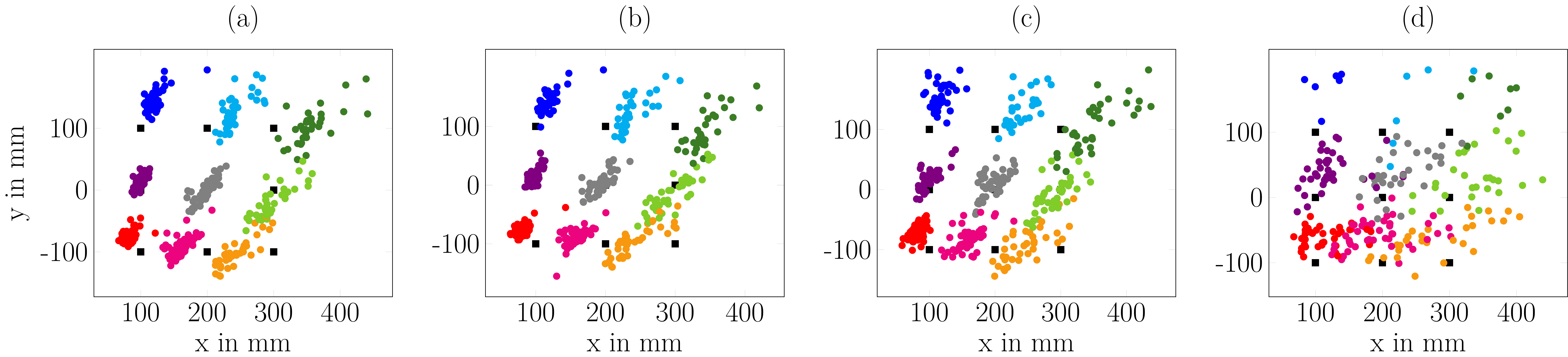}
	\caption{Scatter plot for all resolutions of eye regions of condition \textit{FINGER}. Resolution of eye region is (a) 1000 x 1000,  (b) 500 x500, (c) 250 x 250, (d) 125 x 125. The color marks the aimed positions (x, y): red (100, -100), magenta (200, -100), orange (300, -100), violet (100, 0), gray (200, 0) light green (300, 0), blue (100, 100), light blue (200, 100) and green (300, 100).}
	\label{fig:scatterScenario2}
\end{figure*}
%The scatter plot with the position of all calculated and expected positions is shown in figure \ref{fig:scatter}.

The scatter plot with the position of all calculated and expected positions is shown in Figure \ref{fig:scatterScenario1} for condition \textit{RECT} and in Figure \ref{fig:scatterScenario2} for condition \textit{FINGER}. %\textcolor{red}{During the evaluation we used head mounted marker for the eye detection. The object detection is based on color based threshold algorithm.} %The RMS error for the full resolution image over all points is 31.13 mm (sd=20.79), for 500x500 pixel 20.69 (sd=19.76), for 250x250 pixel 28.48 (sd=35.7) and for 125x125 pixel 47.86 (sd=32.01).
For condition \textit{RECT}, the root mean square error (RMS) for the full resolution image over all points was 40.65 mm (sd=28.13) with a  99.53\% detection rate. For 500x500 px, it was 42.62mm  (sd=30.87, 95.28\% detection rate), for 250x250 pixel, 42.22 mm (sd=30.60, 72.17\% detection rate) and for 125x125 px, 44.53 mm (sd=27.57, 61.08\% detection rate), see Figure \ref{fig:scatterScenario1}.

As we witnessed a sharp decline in location accuracy at a horizontal distance of 30~cm, we also calculated results for closer interaction distances. The RMS error for the full resolution image with an expected horizontal interaction distance x=100 mm and y=0 mm is 16.38 mm (sd=7.79). For a distance range of y from -100 mm and 100 mm at an x distance of 100 mm the mean location accuracy is 21.51 mm (sd=12.22). For an x range of 100 to 200 mm and y in range of -100 mm and 100 mm is the mean location accuracy is 30.48 mm (sd=19.09).

For condition \textit{FINGER}, the RMS error for the full resolution image over all points is 41.11 mm (sd=24.02, 93.17\% detection rate), for 500x500 px, 41.25 mm (sd=23.94, 89.76\% detection rate), for 250x250 px, 45.18 mm (sd=26.29, 86.59\% detection rate) and for 125x125 px 62.90 mm (sd=30.60, 57.32\% detection rate), see Figure \ref{fig:scatterScenario2}. As in the previous scenario, we also looked at closer interaction ranges. The RMS error for the full resolution image with expected x=100 mm and y=0 mm is 16.12 mm (sd=9.06), for x=100 mm and y in range of -100 mm and 100 mm is 33.13 mm (sd=20.41). For x in the range of 100 to 200 mm and y in range of -100 mm and 100 mm the mean location accuracy is 35.90 mm (sd=21.48).

\begin{figure}[!t]
	\centering
	\includegraphics[width=0.7\columnwidth]{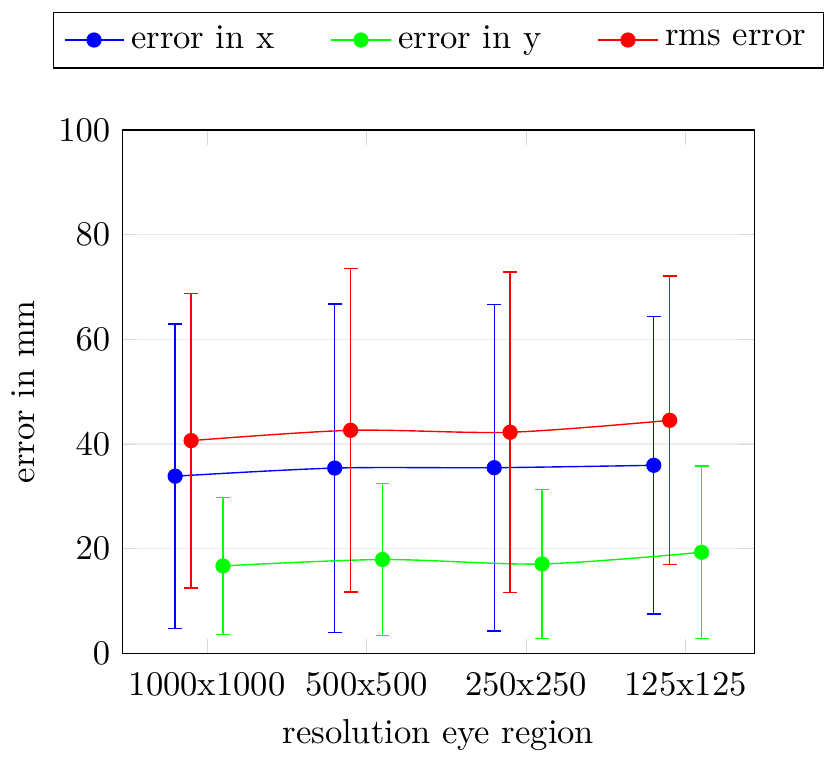}
	\caption{Errors (x, y, RMS) across all resolutions for condition 1.}
	\label{fig:FehlerAllScenario1}
\end{figure}

\begin{figure}[!t]
	\centering
	\includegraphics[width=0.7\columnwidth]{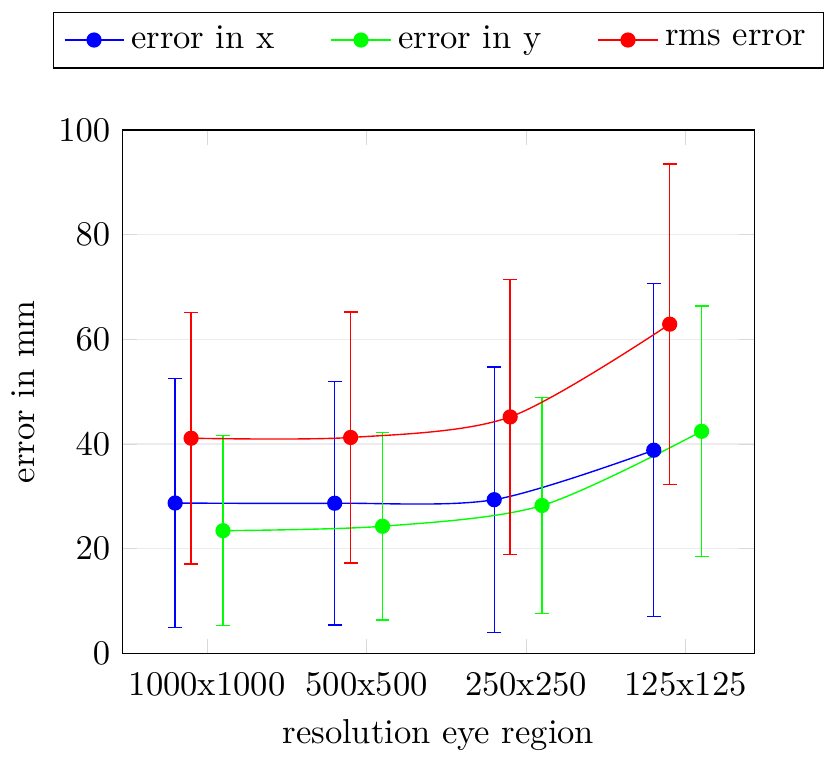}
	\caption{Errors (x, y, RMS) across all resolutions for condition 2.}
	\label{fig:FehlerAllScenario2}
\end{figure}

It is also possible to explain the behavior over the resolutions (see Figure \ref{fig:FehlerAllScenario1} and Figure \ref{fig:FehlerAllScenario2}). At the largest resolution, a pixel corresponds to approximately 0.12\(^{\circ}\). The smaller the resolution, the larger the space that is projected onto a single pixel. The smallest resolution (125x125 px) results in a pixel corresponding to approximately 1\(^{\circ}\). In addition to the direct distance, the calculation of the projection of the red rectangle (for the construction of the plane) leads to considerable fluctuations (similar for the y-direction). %Because of the smaller distances, this has only a small  does not have any effect on the change.

% The same behavior can be seen for the combined error in figure \ref{fig:DistanzFehlerAll}

% \begin{figure}[!t]
% 	\centering
% 	\includegraphics[width=\linewidth]{DistanzFehlerAll.pdf}
% 	\caption{RMS error separated in x and y}
% 	\label{fig:DistanzFehlerAll}
% \end{figure}

The algorithm was executed on a ThinkPad S531 with a Intel Core i7-3537U processor (2 cores, 2.0GHz, 4MB cache) and 10 GB DDR3 RAM. We measured the time the prototype needed to calculate the position of the objects for all tested resolutions. For condition \textit{RECT} and an eye region with resolution of 1000x1000 pixel the mean time to complete the task over all tests %(two persons, each nine position, each 20 times algorithm) 
is 137 milliseconds (sd=41; eye detection 31 ms, limbus detection 35 ms, unwrapping 62 ms, scene analysis 8 ms).  For the resolution with 500x500 pixel average processing time is 68 milliseconds (sd=11; eye detection 27 ms, limbus detection 24 ms, unwrapping 13 ms, scene analysis 2 ms), for the resolution of 250x250 pixel it is 61 milliseconds (sd= 13; eye detection 27 ms, limbus detection 30 ms, unwrapping 3 ms, scene analysis 0.7 ms) and for 125x125 pixel the average time is 63 milliseconds (sd=10; eye detection 26 ms, limbus detection 33 ms, unwrapping 3 ms, scene analysis 0.6 ms). For condition \textit{FINGER} the mean processing time is 287 milliseconds (sd=110; eye detection 31 ms, limbus detection 37 ms, unwrapping 63 ms, scene analysis 154$\pm$100 ms) for the full resolution, for the eye region of 500x500 pixel it is 140 milliseconds (sd=48; eye detection 31 ms, limbus detection 27 ms, unwrapping 14 ms, scene analysis 66 ms), for 250x250 pixel 92 milliseconds (sd=22; eye detection 29 ms, limbus detection 32 ms, unwrapping 3 ms, scene analysis 26 ms) and for 125x125 pixel it is 111 milliseconds (sd=27; eye detection 33 ms, limbus detection 43 ms, unwrapping 4 ms, scene analysis 31 ms). The scene analysis in condition \textit{FINGER} takes longer due to the more complex algorithm. For the detection of the hand and finger an iterative flood-fill algorithm is used. In each iteration the colors are adapted until the hand is detected. The resolution of the image processed depends on the individual steps. For eye detection the image resolution is fixed to 960x614 pixel %(image size of eye resolution of 125x125 pixel)
. For the limbus detection the image (eye region) is scaled to 125x125 pixel. The unwrapping and object detection step work with the original resolution, with exception of the smallest. To improve the success rate of the object detection these images are scaled up to a size of 250x250 pixel for the eye region.
% \(0.0097\pm0.0038\). %\(0.0614\pm0.0237\) seconds, for the resolution of 250 x 250 it is \(0.0141\pm0.0050\) seconds and for 125 x 125 the time is \(0.0097\pm0.0038\).
%\lipsum

\section{Discussion and Conclusion}\label{sec:conclusion}
%\lipsum
% \textcolor{red}{TODO: typische Augenauflößung}
% \textcolor{red}{TODO: verkleinern auf 4k / FullHD}
The evaluation has indicated that, under optimal conditions, in a region of 20 cm next to and 10 cm above or below the center point of a mobile device, positions between two sensed objects (such as a finger) could be distinguished if these are approximately 5 cm apart. Please note, that both conditions took place under optimal laboratory conditions. This was done on purpose to get a best case estimate. Future work should validate these findings in more ecological valid settings. To this end, the scene analysis step would need further adoption to increase robustness to different lighting conditions. 

In both scenarios, the error increases with the x-coordinate (distance between the two objects in x-direction). This behavior can be partly explained due to the curvature of the reflection, which results at lower image resolutions for eye surface regions facing away from the view direction.  %The normal of a sphere can be projected onto a plane. The normal that is parallel to the normal of the plane and intersects it in the origin at the positive distance, is the reference standard. The distance between origin and an intersection of another normal can be calculated with the tangent of the angle between the two normals. Due to the use of the tangent, the distance becomes larger at larger angles. In that the angle is determined from a pixel image, a deviation near the origin leads to a small deviation in the x-direction and the further away this deviation occurs, the more it acts. 
Further, with decreasing resolution of the eye region, the detection rate drops dramatically. For example, the blue marker in condition \textit{RECT} has a size of about 6x6 pixel at full resolution. At other resolutions, the color is already mixed with the surrounding background or iris texture. The second condition has the same problem. Here the width of the index finger of the participants were typically even smaller then the width of the blue marker.

A further deviation comes from the fault-prone object recognition, which, albeit the colored rectangles in condition \textit{RECT}, amplifies errors due to pixel-wise offsets in the position detection. %Since these markings mainly have an influence on the width, the deviations are smaller here. 
For example, at 5 x 10 pixel ground truth size, by detecting 6 x 10 pixels, deviations of 20\% in x-direction and 10\% in y-direction result. With more complex objects (hand) or other environments, the detection error might increase. One way to potentially aid object recognition, is the inclusion of iris texture removal algorithms as a pre-processing step before scene analysis (e.g.,\cite{wang2008separating}). Besides required resolution, a limitation of our pipeline is that the algorithm is not directly optimized for execution on mobile devices.

%\todo[inline]{WHY 10\% in y??}
%Furthermore, only very close points (\(\pm100 mm \)) are shown in the y-direction. This causes the errors in the y-direction to remain constant across the resolutions.% In order for these to work for the small resolutions, the objects had to be marked by hand.

 %Thus, when developing the concept and implementing the prototype, no major attention was paid to how many resources are required. % For a real applications on mobile devices it is necessary that the results of the algorithm are not available after several seconds.

Another limitation may be the currently assumed optimal gaze direction. With head or eye movements present, the objects may be projected towards outer regions of the cornea or even sclera. This in turn can lead to challenging object detection (due to larger overlap with iris texture) and to higher uncertainties in object localization (due to a stronger oval shape of the cornea in image space, which leads to a lower image space resolution). However, under the assumption that users look at the screen while interacting, the deviations from our setup can be small. 

%Another limitation is the poor object detection. In the currently used object recognition, errors already occur in simple test objects (red rectangle and blue square) in the test environment. With more complex objects (hand) or other environments, the detection error might increase. The recognition results might improve if additional iris texture removal algorithms (c.f. \cite{wang2008separating}) are applied prior to the scene analysis step.

Also, since the detection of the limbus is based on a selection of random values, this results in strong fluctuations of the results. The algorithm we employed  \cite{wood2014eyetab} has an accuracy of 7\(^{\circ}\) for the gaze direction, other state of the art approaches (c.f., \cite{zhang2016s}) can achieve an accuracy of 5\(^{\circ}\) and could be applied in the future.

%By adjusting the limbus detection, a higher accuracy can be achieved, which leads to the determination of the position being more accurate.

%A further improvement can be achieved by the adaptation of the object recognition. The change to a more robust object detection would lead to the deviations being reduced. An additional tracking could also help the object detection and improve the accuracy further.

%If the capturing camera is permanently installed with the mobile display, this property can be used for object recognition. A ray, which emanates from the pinhole of the camera, is reflected on the corea, back to the pinhole. For such a reflection, the normal of the surface must be parallel to the ray through the pinhole. This results in the ray passing through the center of the corneal sphere. The pixel, in which the beam intersects the image plane, thus represents the pinhole of the camera in the image. Thus the object recognition can be supported by known size and position of the camera in the terminal.

Finally, the use of the rectangular projection leads to distortions in the image. Another projection or better rotation during unwrapping can also provide for improvement.

To summarize, we work towards extending the input area of mobile devices using front-facing device-centered cameras that capture reflections in the cornea. To this end, we adapted corneal imaging techniques for the use in an around-device interaction pipeline. We studied the feasibility of around-device interaction using corneal reflective imaging based on a high resolution camera but also investigated the performance on lower resolution images. Our results indicate, that under optimal conditions around-device sensing could be performed with a positional resolution of ca. 5 cm. In future work, we want to optimize this approach to work on mobile phones with built-in high resolution (4k, 8k) cameras and increase the robustness in real-world conditions.
\balance

%\section{References format}
%References must be the same font size as other body text.
% REFERENCES FORMAT
% References must be the same font size as other body text.

\bibliographystyle{acm-sigchi}
\bibliography{iss-short}

\begin{thebibliography}{10}

\bibitem{bababekova2011font}
Bababekova, Y., Rosenfield, M., Hue, J.~E., and Huang, R.~R.
\newblock Font size and viewing distance of handheld smart phones.
\newblock {\em OaVS 88}, 7 (2011), 795--797.

\bibitem{backes2009tempest}
Backes, M., Chen, T., D{\"u}rmuth, M., Lensch, H.~P., and Welk, M.
\newblock Tempest in a teapot: Compromising reflections revisited.
\newblock In {\em 30th IEEE SSaP}, IEEE (2009), 315--327.

\bibitem{butler2008sidesight}
Butler, A., Izadi, S., and Hodges, S.
\newblock Sidesight: Multi-"touch" interaction around small devices.
\newblock In {\em Proc. UIST '08}, ACM (2008), 201--204.

\bibitem{buxton2008long}
Buxton, B.
\newblock The long nose of innovation.
\newblock {\em Insight 11\/} (2008), 27.

\bibitem{chan2015cyclops}
Chan, L., Hsieh, C.-H., Chen, Y.-L., Yang, S., Huang, D.-Y., Liang, R.-H., and
  Chen, B.-Y.
\newblock Cyclops: Wearable and single-piece full-body gesture input devices.
\newblock In {\em Proc. CHI '15}, ACM (2015), 3001--3009.

\bibitem{chen2014duet}
Chen, X., Grossman, T., Wigdor, D.~J., and Fitzmaurice, G.
\newblock Duet: exploring joint interactions on a smart phone and a smart
  watch.
\newblock In {\em Proc. CHI '14}, ACM (2014), 159--168.

\bibitem{DeLuca:2013:BAS:2470654.2481330}
De~Luca, A., von Zezschwitz, E., Nguyen, N. D.~H., Maurer, M.-E., Rubegni, E.,
  Scipioni, M.~P., and Langheinrich, M.
\newblock Back-of-device authentication on smartphones.
\newblock In {\em Proc. CHI '13}, ACM (2013), 2389--2398.

\bibitem{esteves2015orbits}
Esteves, A., Velloso, E., Bulling, A., and Gellersen, H.
\newblock Orbits: enabling gaze interaction in smart watches using moving
  targets.
\newblock In {\em Proc. UbiComp/ISWC'15 Adj.}, ACM (2015), 419--422.

\bibitem{Freeman:2014:TUA:2628363.2634215}
Freeman, E., Brewster, S., and Lantz, V.
\newblock Towards usable and acceptable above-device interactions.
\newblock In {\em Proc. MobileHCI '14}, ACM (2014), 459--464.

\bibitem{729537}
Gemperle, F., Kasabach, C., Stivoric, J., Bauer, M., and Martin, R.
\newblock Design for wearability.
\newblock In {\em Wearable Computers, 1998. Digest of Papers. Second
  International Symposium on} (Oct 1998), 116--122.

\bibitem{grubert2015multifi}
Grubert, J., Heinisch, M., Quigley, A., and Schmalstieg, D.
\newblock Multifi: Multi fidelity interaction with displays on and around the
  body.
\newblock In {\em Proc. CHI '15}, ACM (2015), 3933--3942.

\bibitem{grubert2016glasshands}
Grubert, J., Ofek, E., Pahud, M., Kranz, M., and Schmalstieg, D.
\newblock Glasshands: Interaction around unmodified mobile devices using
  sunglasses.
\newblock In {\em Proc. ISS '16}, ACM (2016), 215--224.

\bibitem{harrison2010appropriated}
Harrison, C.
\newblock Appropriated interaction surfaces.
\newblock {\em Computer 43}, 6 (2010), 0086--89.

\bibitem{kawaguchi2005detection}
Kawaguchi, T., Rizon, M., and Hidaka, D.
\newblock Detection of eyes from human faces by hough transform and
  separability filter.
\newblock {\em Electronics and Communications in Japan (Part II: Electronics)
  88}, 5 (2005), 29--39.

\bibitem{Khamis:2016:CDS:2968219.2968342}
Khamis, M., Alt, F., and Bulling, A.
\newblock Challenges and design space of gaze-enabled public displays.
\newblock In {\em Proc. UbiComp/ISWC'16 Adj.}, ACM (2016), 1736--1745.

\bibitem{Khamis:2016:GMA:2851581.2892314}
Khamis, M., Alt, F., Hassib, M., von Zezschwitz, E., Hasholzner, R., and
  Bulling, A.
\newblock Gazetouchpass: Multimodal authentication using gaze and touch on
  mobile devices.
\newblock In {\em Proc. CHI EA '16}, ACM (2016), 2156--2164.

\bibitem{kothari1996detection}
Kothari, R., and Mitchell, J.~L.
\newblock Detection of eye locations in unconstrained visual images.
\newblock In {\em Proc. ICIP '96}, vol.~3, IEEE (1996), 519--522.

\bibitem{kratz2009hoverflow}
Kratz, S., and Rohs, M.
\newblock Hoverflow: exploring around-device interaction with ir distance
  sensors.
\newblock In {\em Proc. MobileHCI '09}, ACM (2009), 42.

\bibitem{nandakumar2016fingerio}
Nandakumar, R., Iyer, V., Tan, D., and Gollakota, S.
\newblock Fingerio: Using active sonar for fine-grained finger tracking.
\newblock In {\em Proc. CHI '16}, ACM (2016), 1515--1525.

\bibitem{nishino2004eyes}
Nishino, K., and Nayar, S.~K.
\newblock Eyes for relighting.
\newblock In {\em ACM TOG}, vol.~23, ACM (2004), 704--711.

\bibitem{nishino2006corneal}
Nishino, K., and Nayar, S.~K.
\newblock Corneal imaging system: Environment from eyes.
\newblock {\em IJCV 70}, 1 (2006), 23--40.

\bibitem{nitschke2013see}
Nitschke, C., Nakazawa, A., and Nishida, T.
\newblock I see what you see: point of gaze estimation from corneal images.
\newblock In {\em Proc. ACPR '13}, IEEE (2013), 298--304.

\bibitem{nitschke2011display}
Nitschke, C., Nakazawa, A., and Takemura, H.
\newblock Display-camera calibration using eye reflections and geometry
  constraints.
\newblock {\em Computer Vision and Image Understanding 115}, 6 (2011),
  835--853.

\bibitem{nitschke2013corneal}
Nitschke, C., Nakazawa, A., and Takemura, H.
\newblock Corneal imaging revisited: An overview of corneal reflection analysis
  and applications.
\newblock {\em IPSJ Transactions on Computer Vision and Applications 5\/}
  (2013), 1--18.

\bibitem{Oakley:2014:IEO:2556288.2557138}
Oakley, I., and Lee, D.
\newblock Interaction on the edge: Offset sensing for small devices.
\newblock In {\em Proceedings of the SIGCHI Conference on Human Factors in
  Computing Systems}, CHI '14, ACM (New York, NY, USA, 2014), 169--178.

\bibitem{radle2014huddlelamp}
R{\"a}dle, R., Jetter, H.-C., Marquardt, N., Reiterer, H., and Rogers, Y.
\newblock Huddlelamp: Spatially-aware mobile displays for ad-hoc
  around-the-table collaboration.
\newblock In {\em Proc. ITS '14}, ACM (2014), 45--54.

\bibitem{Schmidt:2012:CIS:2317956.2318005}
Schmidt, D., Seifert, J., Rukzio, E., and Gellersen, H.
\newblock A cross-device interaction style for mobiles and surfaces.
\newblock In {\em Proc. DIS '12}, ACM (2012), 318--327.

\bibitem{schnieders2010reconstruction}
Schnieders, D., Fu, X., and Wong, K.-Y.~K.
\newblock Reconstruction of display and eyes from a single image.
\newblock In {\em Proc. CVPR '10}, IEEE (2010), 1442--1449.

\bibitem{FatFinger}
Siek, K., Rogers, Y., and Connelly, K.
\newblock Fat finger worries: How older and younger users physically interact
  with pdas.
\newblock In {\em Human-Computer Interaction - INTERACT 2005}, M.~Costabile and
  F.~Paternò, Eds., vol.~3585 of {\em LNCS}. Springer Berlin Heidelberg, 2005,
  267--280.

\bibitem{snyder1997flattening}
Snyder, J.~P.
\newblock {\em Flattening the earth: two thousand years of map projections}.
\newblock University of Chicago Press, 1997.

\bibitem{song2014air}
Song, J., S{\"o}r{\"o}s, G., Pece, F., Fanello, S.~R., Izadi, S., Keskin, C.,
  and Hilliges, O.
\newblock In-air gestures around unmodified mobile devices.
\newblock In {\em Proc. UIST '14}, ACM (2014), 319--329.

\bibitem{wang2008separating}
Wang, H., Lin, S., Ye, X., and Gu, W.
\newblock Separating corneal reflections for illumination estimation.
\newblock {\em Neurocomputing 71}, 10 (2008), 1788--1797.

\bibitem{Wigdor:2007:LTS:1294211.1294259}
Wigdor, D., Forlines, C., Baudisch, P., Barnwell, J., and Shen, C.
\newblock Lucid touch: A see-through mobile device.
\newblock In {\em Proc. UIST '07}, UIST '07, ACM (New York, NY, USA, 2007),
  269--278.

\bibitem{wood20163d}
Wood, E., Baltru{\v{s}}aitis, T., Morency, L.-P., Robinson, P., and Bulling, A.
\newblock A 3d morphable eye region model for gaze estimation.
\newblock In {\em ECCV}, Springer (2016), 297--313.

\bibitem{wood2014eyetab}
Wood, E., and Bulling, A.
\newblock Eyetab: Model-based gaze estimation on unmodified tablet computers.
\newblock In {\em Proc. ETRA '14}, ACM (2014), 207--210.

\bibitem{xiao2014toffee}
Xiao, R., Lew, G., Marsanico, J., Hariharan, D., Hudson, S., and Harrison, C.
\newblock Toffee: enabling ad hoc, around-device interaction with acoustic
  time-of-arrival correlation.
\newblock In {\em Proc. MobileHCI '14}, ACM (2014), 67--76.

\bibitem{yang2013surround}
Yang, X.-D., Hasan, K., Bruce, N., and Irani, P.
\newblock Surround-see: enabling peripheral vision on smartphones during active
  use.
\newblock In {\em Proceedings of the 26th annual ACM symposium on User
  interface software and technology}, ACM (2013), 291--300.

\bibitem{zhang2016s}
Zhang, X., Sugano, Y., Fritz, M., and Bulling, A.
\newblock It's written all over your face: Full-face appearance-based gaze
  estimation.
\newblock {\em arXiv preprint arXiv:1611.08860\/} (2016).

\bibitem{Zhao:2014:SDI:2642918.2647380}
Zhao, C., Chen, K.-Y., Aumi, M. T.~I., Patel, S., and Reynolds, M.~S.
\newblock Sideswipe: Detecting in-air gestures around mobile devices using
  actual gsm signal.
\newblock In {\em Proc. UIST '14}, ACM (New York, NY, USA, 2014), 527--534.

\end{thebibliography}
\end{document}